\documentclass[a4paper,12pt]{article}
\usepackage{verbatim}
\usepackage{listings}
\usepackage{upquote}
\input{lst-color.sty}
\usepackage{graphicx}
\usepackage[lofdepth,lotdepth]{subfig}
\usepackage[labelfont=bf]{caption}
\usepackage{float}
\usepackage{amsmath}
\usepackage{amssymb} 
\usepackage{amsfonts}
\usepackage{slashed}
\usepackage{stmaryrd}
\usepackage[utf8]{inputenc}     
\usepackage[T1]{fontenc} 
\usepackage{multirow}
\usepackage[english]{babel}     
\usepackage{xcolor} 
\usepackage{tikz-feynman}
\definecolor{linkcolor}{rgb}{0,0,0.6} 
\usepackage{hyperref}
\usepackage{amssymb}
\usepackage{soul}
\usepackage{geometry}
\usepackage{latexsym}
\usepackage{cite}
\usepackage{latexsym}
\usepackage{framed}
\usepackage{cite}
\usepackage{soul}

\newcommand{\marty}{\emph{MARTY}}
\newcommand{\csl}{\emph{CSL}}
\newcommand{\grafed}{\emph{GRAFED}}
\newcommand{\martyfull}{Modern ARtificial Theoretical phYsicist}
\newcommand{\op}{\hat{\mathcal{O}}}

\begin{document}
\hfill {\tt  CERN-TH-2020-187} 

\def\thefootnote{\fnsymbol{footnote}}

\begin{center}
\raisebox{-0.365cm}{\includegraphics[width=0.24\linewidth]{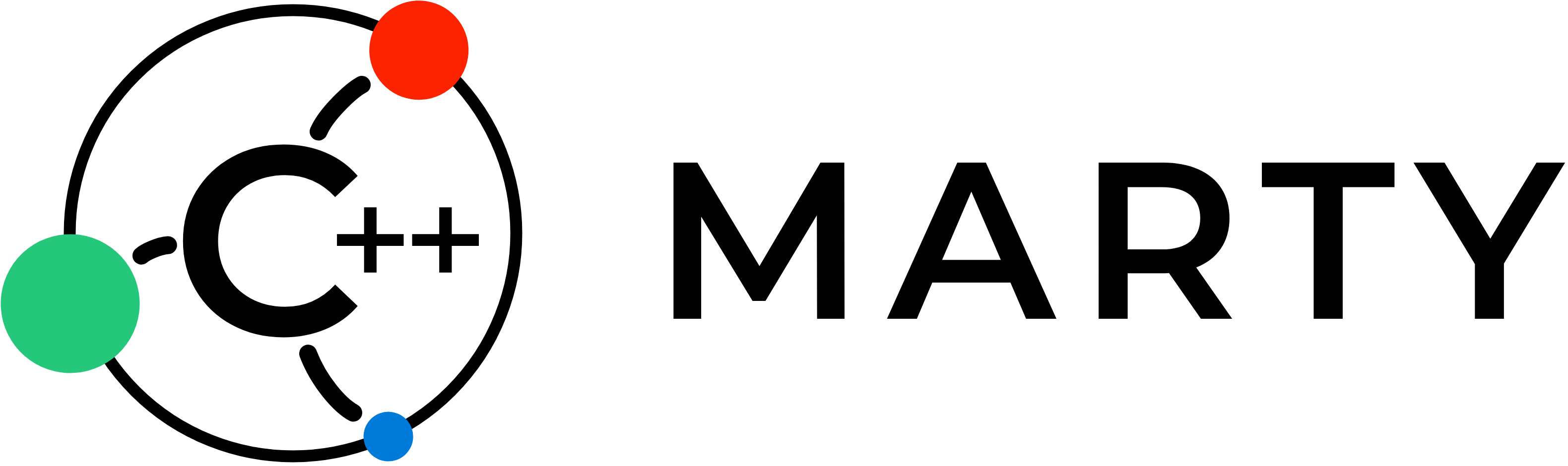}}{\Large\bf{ - \martyfull}\\
\vspace{0.3cm}
A C++ framework automating theoretical calculations Beyond the Standard Model}

\setlength{\textwidth}{11cm}
                    
\vspace{2.cm}
{\large\bf  
G.~Uhlrich$^{a,}$\footnote{Corresponding author, email: g.uhlrich@ipnl.in2p3.fr},
F.~Mahmoudi$^{a,b,}$\footnote{Also Institut Universitaire de France, 103 boulevard Saint-Michel, 75005 Paris, France},
A.~Arbey$^{a,b,\dagger}$
}
 
\vspace{1.cm}
{\em $^a$Universit\'e de Lyon, Universit\'e Claude Bernard Lyon 1, CNRS/IN2P3, \\
Institut de Physique des 2 Infinis de Lyon, UMR 5822, F-69622, Villeurbanne, France}\\[0.2cm]
{\em $^b$Theoretical Physics Department, CERN, CH-1211 Geneva 23, Switzerland} 

\end{center}

\renewcommand{\thefootnote}{\arabic{footnote}}
\setcounter{footnote}{0}

\vspace{1.cm}
\thispagestyle{empty}
\centerline{\bf ABSTRACT}
\vspace{0.5cm}
Studies Beyond the Standard Model (BSM) will become more and more important in the near future with the rapidly increasing amount of data from different experiments around the world. The full study of BSM models is in general an extremely time-consuming task involving long and difficult calculations. It is in practice not possible to do exhaustive predictions in these models by hand, in particular if one wants to perform a statistical comparison with data and the SM.

Here we present MARTY (Modern ARtificial Theoretical phYsicist), a new C++ framework that fully automates calculations from the Lagrangian to physical quantities such as amplitudes or cross-sections. This framework can fully simplify, automatically and symbolically, physical quantities in a very large variety of models. MARTY can also compute Wilson coefficients in effective theories. This will considerably facilitate BSM studies in flavour physics. 

Contrary to the existing public codes in this field MARTY aims at providing a unique, free, open-source, powerful and user-friendly tool for high-energy physicists studying predictive BSM models, in effective or full theories up to the one-loop level, which does not rely on any external package. With a few lines of code one can gather final expressions that may be evaluated numerically for statistical analysis. Features such as automatic generation and edition of Feynman diagrams, comprehensive manuals and documentation, clear and easy to handle user interface are amongst notable features of MARTY.

\clearpage

\section{Introduction}

Calculations Beyond the Standard Model (BSM) have been a challenge for a long time, especially at the one-loop level. Transition amplitudes, differential cross-sections, and Wilson coefficients need to be evaluated in order to obtain the values for physical observables. Comparing them to the SM predictions and experimental results provides means to discriminate BSM models.

The one-loop order is often needed, as some relevant processes only appear at this order. This is the case of Flavour Changing Neutral Currents (FCNC) in flavour physics \cite{buras}, that require one-loop values for Wilson coefficients. At the one-loop level, calculations cannot be done numerically as the very large number of terms together with renormalization processes require symbolic computations similar to what can be done by hand.

Currently, symbolic calculations at the one-loop level are done mainly using
Mathematica \cite{mathematica}, a commercial and closed computer algebra system. Several packages based on Mathematica implement high-energy physics calculations. FeynRules \cite{feynrules} computes Feynman rules from a BSM Lagrangian, from which FormCalc \cite{formcalc} can derive tree-level and one-loop quantities such as transition amplitudes or differential cross-sections. Packages such as FormFlavor \cite{formflavor} or FlavorKit \cite{sarah, flavorkit} calculate Wilson coefficients using the FormCalc machinery.

Many efforts have been employed to design independent computer algebra systems for high-energy physics. GiNaC \cite{ginac} is a C++ library for symbolic computations written for that purpose. LanHEP \cite{lanhep}, CompHEP \cite{comphep} and CalcHEP \cite{calchep} automate together calculations from the Lagrangian using their own symbolic computation framework. However, Mathematica-based packages are still the only ones able to provide one-loop calculations and Wilson coefficients for BSM.

We introduce \marty, a modern and user-friendly solution to this issue. It provides for the first time a unique, free and open-source code fully written in modern C++ (2017 standard), implementing all the theoretical BSM machinery up to the one-loop order. Amplitudes, cross-sections and Wilson coefficients up to dimension 5 operators can be computed automatically. Support for dimension 6 operators will come in a future version. \marty\ does not depend on any other framework, as it includes its own C++ Symbolic computation Library called \emph{CSL} to manipulate mathematical expressions. It also comes with \grafed, a desktop application displaying Feynman diagrams on screen.

The reader can find the code, manuals, the documentation and more information on \url{https://marty.in2p3.fr}.
\section{Code overview}
\begin{figure}[h]
    \centering
    \includegraphics[width=0.85\linewidth]{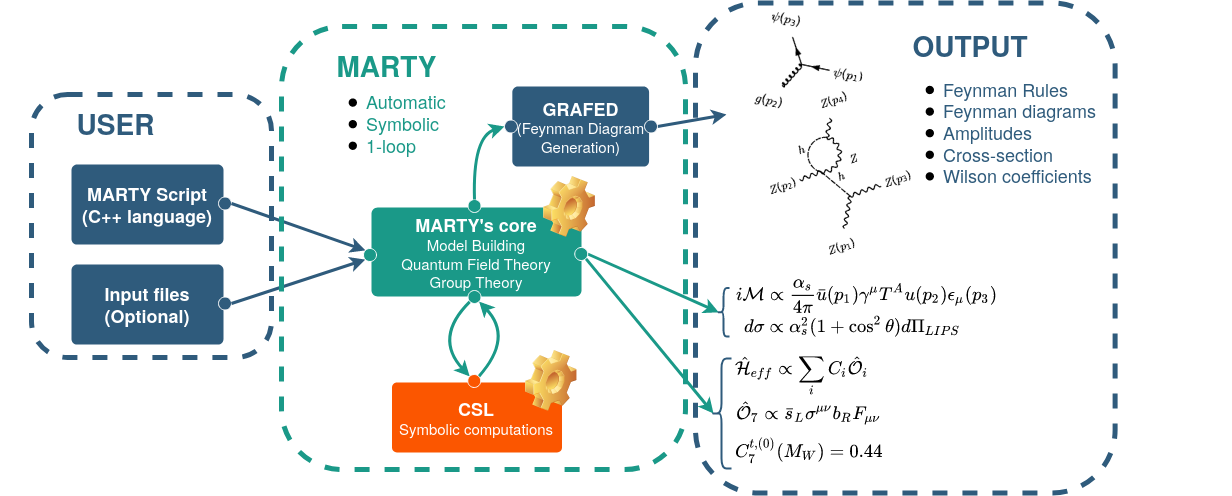}
    \caption{Schematic design of \marty, including the three sub-modules: \marty\ (physics part), \csl\ (symbolic manipulation) and \grafed\ (Feynman diagram generation).}
    \label{fig:design}
\end{figure}

\marty\ is organised in several modules. They have been logically separated to ensure the code to be modular and general. The results are three modules: the physics core of \marty, and two other modules which are both fully independent from each other and can be used as standalone, namely the computer algebra system of \marty(\csl) and the application drawing Feynman diagrams (\grafed). A schematic view of \marty's design is presented in figure \ref{fig:design}.

\subsection{\marty\ (physics core)}
This is the main part of the code, which contains all physics calculations, conventions, models, simplifications, etc. It uses its own \csl\ module as a mathematical backend to perform calculations specific to high-energy physics and the \grafed\ module to draw Feynman diagrams. It also contains all group theory implementations, and model building features. Amplitudes, differential partonic cross-sections and Wilson coefficient may be calculated at tree-level or at the one-loop order. All calculations are automatic, symbolic, and can be performed in a very large variety of models as detailed in section \ref{capabilities}.

\subsection{\csl\ (computer algebra system)}
This module does not know anything about physics, is logically separated from the physics part and can be used independently. It is a C++ Symbolic computation Library allowing us to handle mathematical expressions, tensors and simplifications needed to perform high-energy physics calculations. It is not as comprehensive as a standard computer algebra system like Mathematica because many features were not required for particle physics. It may however be extended in that direction if needed.

\subsection{\grafed\ (Feynman diagram generation and edition)}
When doing calculations in particle physics, it is often convenient to visualize what the code is doing, and possibly include diagrams in publications. \grafed\ was developed for this purpose and is also fully independent of the other modules of \marty. It has three major features:
\begin{itemize}
    \item An algorithm that finds an optimal way to place nodes in a 2D space to display Feynman diagrams. This algorithm is fully general (with no limit in the diagram size or number of loops) and automated. This allows one to quickly draw all diagrams for a particular process, without asking anything from the user and independently of the diagram topologies.
    \item A Graphical User Interface (GUI) that displays the generated diagrams. When asked, \marty\ will run \grafed\ with all the diagrams of a particular process. These diagrams appear then in the GUI, and may be exported (as png files or LATEX\ codes for the tikz-feynman package) directly to be included in a publication for example. 
    \item The possibility to edit or create diagrams from scratch. Diagrams generated automatically by \grafed\ are rather neat, but there is the possibility to edit, graphically, any aspects of the diagram (nodes, edges, labels, layout, etc) very easily. One can also create diagrams from scratch using \grafed\ independently of \marty.
\end{itemize}

\section{\marty\ design philosophy}

The design of \marty\ is guided by strong principles ensuring a final result corresponding to programming standards. First, the general principles unrelated to physics are:
\begin{itemize}
    \item \textbf{Independence.} \marty\ is written from scratch and is thus fully independent of any other framework. As such, there is no limit in what can be implemented in the code. Developers have a full control on any aspects of \marty, to modify or extend its capabilities.
    \item \textbf{User-friendliness.} The code must be easy to use. The fact that it is written in C++ is a supplementary challenge in that purpose, but a modern knowledge of this language provides freedom for the user-interface. We think that this objective is fulfilled, since the normal usage of \marty\ does not require any particular C++ knowledge and would be similar in many languages, including Mathematica.
    \item \textbf{Modularity.} \marty\ is built as modular as possible. This means that unnecessary logical connections between different parts of the code are avoided. This is an important advantage for maintainability, since replacing or correcting a part of the code will become simpler. 
    \item \textbf{Readability.} It is important for a code to be easily understandable by everyone, and in particular by a user willing to further develop the code. Strict coding conventions, clear naming for files / functions / variables and clever separation of different logical units make \marty\ easy to understand considering its large size.
    \item \textbf{Performance.} C++ and python were the two main languages possible for \marty\ as they are well-known in the high energy physics community. The choice of the language, C++, is related to performance reasons. A C++ code will run much faster in average than python for this type of code. 
\end{itemize}

Concerning physics aspects, \marty\ has been written with the following aims in mind:
\begin{itemize}
    \item \textbf{Generality.} \marty\ is designed to be as general as possible in the models it can handle, algebraic simplifications it can do, and calculations automated with it. A high level of generality has already been reached, and further developments will continue to focus on this aspect. In particular there is no hard-coding because \marty\ is expected to be extended even further in its future developments.
    \item \textbf{Model independent calculations.} In order to have an easy-to-use code for high energy physics, the computations have to be done in a model-independent way. The same code computing a given quantity should work for all models. Studying new models would hence imply only to write the Lagrangian or Feynman rules associated to it, and then using the same scripts to calculate the same quantities in the given model.
    \item \textbf{One-loop level automated calculations.} Calculations in BSM phenomenology often require at least one-loop level quantities. Many processes are trivial at tree-level but higher order corrections can be important from a phenomenological point of view when studying BSM models, as it is the case for instance for FCNC decays in flavour physics. The one-loop level being significantly more difficult to calculate by hand than the tree-level, it is important to automate its calculation.
\end{itemize}

The efforts made to respect this philosophy will be useful in \marty's future developments as well. A code as general and independent as \marty\ could benefit from a community effort to be maintained and developed and we think that the way it is written would allow for such a collaborative work.

\section{Installation and usage}

\marty\ is available for download from its website: \url{https://marty.in2p3.fr} where one can find the manuals, documentation, publications and more information.

\subsection{Installation}

\marty\ is open-source, GPL3 licenced and written in C++. It can be installed on linux (Ubuntu/Debian) via the following commands\footnote{Administrator privileges (sudo) will be needed to install \marty\ and its dependencies in standard locations (such as /usr/include and /usr/lib).}:
\begin{framed}
\begin{lstlisting}[language=bash, keywordstyle=\color{black}\bfseries]
    $ cd MARTY
    $ ./setup.sh
    $ make
    $ make install
\end{lstlisting}
\end{framed}
Installation instructions for other Linux distributions and other Operating Systems (Mac-OS, Windows) can be found on the website.

\subsection{Usage (scripting)}

As a C++ framework, \marty\ can be used in a C++ program after writing a few lines to include it, provided it is installed on the computer\footnote{For this to work properly, one needs either to install \marty\ in the standard location (by default, /usr) or make sure that the paths to access include files in C++ (CPATH environment variable on Ubuntu), libraries (LIBRARY\_PATH and LD\_LIBRARY\_PATH environment variable on Ubuntu) and binaries (PATH) contain the installation path of \marty.}:
\begin{framed}
\begin{lstlisting}[language=C++]
    #include <marty.h>
    
    using namespace mty;
    using namespace csl;
    using namespace std;
\end{lstlisting}
\end{framed}
The three lines of \lstinline!using namespace! are not necessary, but will allow us to omit prefixes \lstinline!mty::! (\marty) \lstinline!csl::! (\csl) and \lstinline!std::! (C++ standard library) in front of objects and functions. The main function containing the program can now be written:
\begin{framed}
\begin{lstlisting}[language=C++]
    int main() {
        // Your program
        return 0;
    }
\end{lstlisting}
\end{framed}

\subsection{Usage (compilation)}
In C++, source files need to be compiled before being run. \marty\ uses the C++17 standard that appears in compiler options. To compile a source file \lstinline!main.cpp! into an executable \lstinline!main.x! the following commands are necessary\footnote{\lstinline!g++! may be replaced by another C compiler, \lstinline!clang! for example.}:
\begin{framed}
\begin{lstlisting}[language=bash, keywordstyle=\color{black}\bfseries]
    $ g++ -std=c++17 -c main.cpp -o main.o
    $ g++ -std=c++17 -o main.x main.o -lmarty
    $ ./main.x
\end{lstlisting}
\end{framed}

\subsection{Dependencies}

\marty\ has been written from scratch. Thus it has no dependency before the numerical evaluation of symbolic results. It contains in particular its own computer algebra system, \csl\ (C++ Symbolic computation Library), as a separate module. Therefore for the physics calculations from the Lagrangian to a final one-loop result simplified as much as possible, \marty\ uses nothing but its own code and the C++ standard library.

For the numerical evaluation of results, there are two dependencies. The first, LoopTools \cite{formcalc}, provides numerical values for scalar integrals arising at the one-loop level. Momentum integrals have always the same form, like the following 3-point function integral
\begin{equation}
    I_3 \equiv\int \frac{d^4q}{i\pi ^2}\frac{p_1^\mu p_2^\nu}{(q^2 - m_0^2)((q+p_1)^2-m_1^2)((q+p_2)^2-m_2)^2}.
\end{equation}
A way to treat this kind of integrals (including regularization) is to decompose the result in different possible Lorentz structures, each having a scalar factor in front, that is calculable with standard prescriptions \cite{one-loop}. The decomposition for the 3-point function $I_3$ is the following 
\begin{equation}
 I_3 \equiv C_{00}g^{\mu\nu} + C_{11}p_1^\mu p_1^\nu + C_{12}(p_1^\mu p_2^\nu + p_2^\mu p_1^\nu)+ C_{22}p_2^\mu p_2^\nu,
\end{equation}
with $C_{ij}$ scalar form factors depending on masses and squared momenta in the loop. The decomposition is done analytically by \marty, and the evaluation uses LoopTools functions to determine the values of $C_{ij}$.

The second dependency for the numerical evaluation is GSL \cite{gsl}, a well-known numerical library for C and C++. For complicated models with non trivial mixings (such as supersymmetric models), one has to diagonalize mass matrices to obtain the mass spectrum and mixings of the theory. For example from a non-diagonal squared-mass matrix $M^2$ of $\Phi$, one calculates the eigenvector $\Phi ^\prime$ that diagonalizes the matrix to $M^2_D$
\begin{align}
    &\Phi  = U\Phi^\prime,\\
    &M_D^2 = U^\dagger M^2U.
\end{align}
The symbolic result of \marty\ is fully general (unless specified otherwise by the user) and uses generic symbols for all masses and mixings (matrices $M_D^2$ and $U$ in the example). For the numerical evaluation, input parameters must be given by the user. The diagonalization is then performed to get the spectrum and mixings, and to calculate the final results. The numerical diagonalization is performed using GSL.

Finally, \grafed\ has a Graphical User Interface (GUI) that uses Qt \cite{qt}. Qt is a C++ framework allowing to build desktop applications fairly easily and is free and open-source with a GPL licence.

\section{\marty's capabilities}
\label{capabilities}
This section presents in detail the calculations that can be performed with \marty, the possible models, and the outputs that the code returns to the user.

\subsection{Model building}
\label{sec:model}
First it is important to have a clear view of the BSM models that \marty\ can handle. A model lies in a 4-dimensional Minkowski space-time and is defined by:
\begin{itemize}
    \item \textbf{A gauge group.} The gauge group may be any combination of Semi-simple Lie Groups\footnote{Strictly speaking, groups that have a semi-simple Lie algebra.} ($U(1)$, $SU(N)$, $SO(N)$, $Sp(N)$, $E_6$, $E_7$, $E_8$, $F_4$, $G_2$). The unbroken Standard Model gauge $SU(3)_c\times SU(2)_L\times U(1)_Y$ is an example of such a combination.
    \item \textbf{A particle content.} Each particle is an irreducible representation of the gauge group, i.e. an irreducible representation of each group composing the gauge. A particle may have a spin 0, 1/2 (Weyl, Dirac or Majorana), or 1. All gauge couplings are introduced automatically by \marty\ without any help from the user.
    \item \textbf{Additional couplings.} The user can add any interaction term in the Lagrangian. \marty\ simply checks that combining the unbroken gauge representations of the interacting particles gives indeed a trivial representation\footnote{If it does not, this is the sign of an obvious gauge violating term. If it does, the term may still violate gauge symmetry but it is more difficult to test automatically.}. 
\end{itemize}

There are two ways to build a model in \marty. The first one is the most straightforward way but also the most complicated one. It consists in giving explicitly the full Lagrangian to \marty. Few terms in general are provided by unbroken gauge couplings, in particular when one studies a phenomenological model extending the SM. In the SM, there are about 100 terms to write by hand coming from the symmetry breaking. In the Minimal Supersymmetric extension of the Standard Model (MSSM), several thousands. It is possible to do it but one has to be very careful on every convention, sign and factor in front of each term. A small error can lead to wrong results due to interference between different diagrams for a given process. 

The second option is to define a high energy Lagrangian with all symmetries preserved, and give \marty\ prescriptions to break it. The initial Lagrangian has much less interaction terms and is simpler to write. Based on correct prescriptions (gauge, flavour symmetry breaking, replacements, renaming, etc) \marty\ will basically re-derive the final Lagrangian for the user. This solution will not necessarily be the easiest one depending on the model but is certainly a practical option. It is in particular the way chosen to build the MSSM in \marty.

In the following we present a sample code building a $SU(2)_L$ gauge with one quark in the doublet representation
\begin{equation}
    Q_L = \left(\begin{array}{c c}
         u_L&  d_L 
    \end{array} \right),
\end{equation}broken by \marty\ with a single instruction. We also ask the code to rename the broken fermions $Q_1$ and $Q_2$ to $u$ and $d$ which corresponds to standard conventions. With 3 lines of breaking prescriptions \marty\ derives the 17 interaction terms (including vector-ghost interactions\footnote{Gauge fixing terms for ghosts are not taken into account here.}) between the final 8 particles in the model.
\begin{framed}
\begin{lstlisting}
    Model model;
    model.addGaugedGroup(group::Type::SU, "L", 2); // Adding a SU(2)
    model.init();

    Particle Q = weylfermion_s("Q", model, Chirality::Left); 
    Q->setGroupRep("L", {1}); // Doublet rep of SU(2), Dinkin label 1
    model.addParticle(Q);

    cout << "Before symmetry breaking : " << endl;
    cout << model << endl;

    model.breakGaugeSymmetry("L");
    model.renameParticle("Q_1", "u"); // Broken fields are named with _1, _2
    model.renameParticle("Q_2", "d"); // etc. by convention

    cout << "After symmetry breaking : " << endl;
    cout << model << endl;
\end{lstlisting}
\end{framed}
For a more evolved model one needs more instructions to specify every conventions, but this method is still very practical and more efficient than giving the full Lagrangian by hand. For a gauge symmetry breaking, explicit expressions of gauge generators ($T^A_{ij}$, $f^{ABC}$) must be known and \marty\ will not necessarily know them. For $SU(2)$ and $SU(3)$ SM gauge terms, the whole procedure is automated, but for other groups the user may have to define expressions for generators.

\marty\ also contains built-in models that can be used directly for calculations: Scalar $\phi^3$ theory, Scalar QED, QED, QCD, Electroweak model, Standard Model, 2 Higgs Doublet Models and Minimal Supersymmetric Standard Models (unconstrained and phenomenological).

\subsection{Amplitudes}
\label{sec:ampl}
Transition amplitudes from an initial state to a final state noted $i\mathcal{M}(i\to f)$ are the basic quantities that \marty\ is able to calculate. It uses the Lagrangian exponentiation as well as the Wick's theorem \cite{Wick} to find all possible diagrams and derive their corresponding expressions. This step is fully general and has no limit in the diagram complexity or in the number of external legs. Amplitudes are then used to calculate differential cross-sections or to derive Wilson coefficients.

Once an analytical expression for a given diagram has been found, it needs to be simplified in several ways in order to obtain a numerical evaluation of the result. Simplification steps done by \marty\ are the following:

    \paragraph{Dirac algebra simplification} This includes calculation of traces in the Dirac space and simplifications in $\gamma-$matrix products, for particles of spin 1/2 \cite{Schwartz}.
    \paragraph{Group algebra simplification} Similarly to $\gamma-$matrices, algebra generators have to be simplified into amplitudes. Projection operators are used \cite{Cvitanovic} and traces are calculated in all semi-simple groups \cite{trace}. The remaining colour structures that cannot be simplified are stored and factored from the rest of the amplitude, in dedicated abbreviations. For standard gauge groups and in particular for fundamental representations all possible terms will be simplified automatically\footnote{Group simplifications are easy to implement in \marty\ in case there are missing ones. The complicated part is to determine theoretically the simplification rules to apply.}.
    \paragraph{Minkowski Index contraction} Minkowski indices are expanded and contracted as much as possible in D-dimensions to perform Dimensional Regularization (DREG) at the one-loop order. As in D-dimension $g^\mu_\mu = D$, one has to expand the whole diagram to gather all factors of $D$.
    \paragraph{Reduction of one-loop momentum integrals} A momentum integral at one-loop can be decomposed on the basis of scalar form factors \cite{one-loop}. These form factors depend on masses and momenta and can be provided by LoopTools \cite{formcalc} up to the rank 4 5-point functions, i.e. loops with five external legs and four momenta in the numerator. This is the actual limit for fully-simplified one-loop quantities in \marty.
    \paragraph{Dimensional Regularization} The form factors coming from one-loop integrals can have a divergent part that is regularized by taking the dimension $D = 4 - 2\epsilon$. In this case, integrals take the form 
    \begin{equation}
        I\approx \frac{a}{\epsilon} + b + \mathcal{O}(\epsilon).
    \end{equation}Factors of $D$ coming from Minkowski index contractions must then be kept to determine the local terms they generate when they are multiplied by a divergent integral \cite{localTerms, localTerms2}. For the scalar one-point function for example, we get the finite part
    \begin{equation}
        \text{Finite}(DA_0(m^2)) = \text{Finite}((4-2\epsilon)A_0(m^2)) = -2m^2 + 4\cdot\text{Finite}(A_0(m^2)).
    \end{equation}
    \paragraph{Equations of motion} For spin 1 particles, the equation of motion is simply 
    \begin{equation}
        \epsilon _\mu(p)p^\mu=0,
    \end{equation} where $\epsilon(p)$ is the polarization 4-vector of the boson. For spin 1/2 particles, the Dirac equation is applied. It reads
    \begin{align}
    &\slashed{p}u(p) = mu(p)\quad\text{for particles,}\\
    &\bar{v}(p)\slashed{p} = -m\bar{v}(p)\quad\text{for anti-particles.}
    \end{align}
    \paragraph{Factorization} Results are partially factored to compactify at most the final expressions. In particular, factorization by masses and momenta are performed.
    \paragraph{Abbreviation} Abbreviations are introduced automatically by \marty\ to lighten expressions and gain in execution time. All abbreviations used can be displayed by typing \lstinline!DisplayAbbreviations()!.
\newline

Using the toy model presented in section \ref{sec:model}, the calculation of an amplitude in \marty\ is very simple. One has to give the order (\lstinline!mty::Order::TreeLevel! here), the model, and the field insertions. Let us consider the process $u\bar{u}\to d\bar{d}$ at tree-level, the following instruction is needed to run the computation:
\begin{framed}
\begin{lstlisting}
    auto res = model.computeAmplitude(
        Order::TreeLevel,
        {Incoming("u"), Incoming(AntiPart("u")),
         Outgoing("d"), Outgoing(AntiPart("d"))}
        );
\end{lstlisting}
\end{framed}
The code is in C++, therefore variable types must be declared. However since C++11, the compiler does not need the user to specify the type returned by a function anymore. By typing \lstinline!auto! the compiler is told to find the exact type itself. Then, \lstinline!res.expressions! contains the different terms of the amplitude and \lstinline!res.diagrams! the Feynman diagrams. To display expressions in standard output, show the diagrams in \grafed and display the abbreviations introduced by \marty\ the following three lines are necessary respectively:
\begin{framed}
\begin{lstlisting}
    Display(res);
    Show(res);
    DisplayAbbreviations();
\end{lstlisting}
\end{framed}

\begin{figure}[!h]
    \centering
    \includegraphics[width=\linewidth]{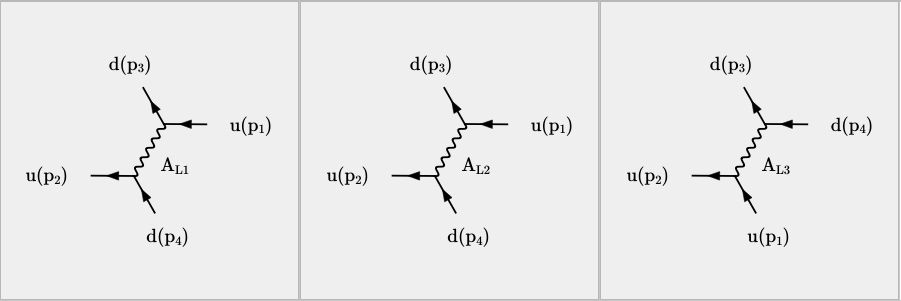}
    \caption{Screenshot of \grafed\ displaying the three Feynman diagrams for the process $u\bar{u}\to d\bar{d}$ at the tree-level in the toy model defined in section \ref{sec:model}. The tree diagrams correspond to the exchange of each broken vector from the $SU(2)_L$ initial symmetry, noted $A_{Li}$ by default in \marty.}
    \label{fig:shotgrafed}
\end{figure}

\marty's output is presented in the following, displaying diagrams in \grafed\ as the screenshot in figure \ref{fig:shotgrafed} shows. Only the index IDs have been changed to lighten the output.
\begin{framed}
\begin{lstlisting}
    0  :  1/8*Ab_0001*EXT_{k,\%eps_1,i,\%del_1,l,\%del_2,j,\%eps_2}
        *gamma_{+\%sigma,\%del_2,\%eps_1}*gamma_{\%sigma,\%del_1,\%eps_2}
    1  :  1/8*Ab_0001*EXT_{k,\%eps_1,i,\%del_1,l,\%del_2,j,\%eps_2}
        *gamma_{+\%sigma,\%del_2,\%eps_1}*gamma_{\%sigma,\%del_1,\%eps_2}
    2  :  -1/8*Ab_0002*EXT_{k,\%eps_1,i,\%del_1,l,\%del_2,5,\%eps_2}
        *gamma_{+\%sigma,\%del_1,\%eps_1}*gamma_{\%sigma,\%del_2,\%eps_2}

    Total : 3 particle amplitudes.
    4 abbreviations:
    Ab = g_L^2
    EXT = d_{k,del}(p_4)*d_{i,gam}(p_3)^(*)*u_{l,beta}(p_2)^(*)*u_{j,alpha}(p_1)
    Ab_0001 = i*g_L^2/s_13
    Ab_0002 = i*g_L^2/s_12
\end{lstlisting}
\end{framed}
This expression can then be read off form \marty's output, keeping a separate term for each diagram:
\begin{equation}
\begin{split}
    i\mathcal{M} = \frac{ig_L^2}{8}\bigg(
        &\frac{1}{s_{13}}\bar{d}(p_4)\gamma ^\mu u(p_1)\bar{u}(p_2)\gamma _\mu d(p_3)\\
        +&\frac{1}{s_{13}}\bar{d}(p_4)\gamma ^\mu u(p_1)\bar{u}(p_2)\gamma _\mu d(p_3)\\
        -&\frac{1}{s_{12}}\bar{u}(p_2)\gamma ^\mu u(p_1)\bar{d}(p_4)\gamma _\mu d(p_3)\bigg),
\end{split}
\end{equation}
with $s_{ij} \equiv p_i\cdot p_j.$

\subsection{Cross-sections}
\label{sec:xsec}
Cross-sections are the main observables used in collider physics. They are directly proportional to the number of events observed in the various detectors. \marty\ does not compute directly the cross-sections but calculates the complicated theoretical part namely the squared amplitudes. For incoming particles $\{I\}$ of spins $\{j_I\}$ and outgoing particles $\{O\}$ of spins $\{j_O\}$ the squared amplitude is (as a function of the amplitude $i\mathcal{M}$ that depends on the particle spins)
\begin{equation}
\label{eq:xsec}
   \frac{1}{\prod _I d_I} \sum _{\{j_I\},\{j_O\}}|\mathcal{M}|^2,
\end{equation}
with $d_I$ the spin dimension of the incoming particle $I$ taking into account massless effects for spin 1 particles. This quantity is averaged (summed) over the spin dimensions of incoming (outgoing) particles. Calculating the squared amplitudes implies the calculation of traces in Dirac and colour spaces (group algebra) that \marty\ computes automatically. The result is a scalar depending on momenta and masses of particles in the process. The differential cross-section has always the same form for a given process of amplitude $i\mathcal{M}$
\begin{equation}
     d\sigma \equiv K(p_i, m_i)\cdot \frac{1}{\prod _I d_I} \sum _{\{j_I\},\{j_O\}}|\mathcal{M}|^2d\Pi _{LIPS},
\end{equation}
with $K(p_i, m_i)$ a factor coming from kinematics, and $d\Pi _{LIPS}$ the Lorentz Invariant Phase Space. Once the amplitude squared has been calculated and simplified, no more computer algebra system is needed to pursue the calculation. This is the quantity that \marty\ can compute automatically.

Considering the toy model of section \ref{sec:model}, calculating the squared amplitude is very simple. The user must first calculate the amplitude, and simply square it with \marty. The average over incoming spins is done by \marty, i.e. the returned quantity corresponds to equation \ref{eq:xsec}. After calculating this quantity with \marty, the user has again one single line to write:
\begin{framed}
\begin{lstlisting}
    Expr squared_ampl = model.computeSquaredAmplitude(res);
    cout << "<|M|^2> = " << squared_ampl << endl;
    DisplayAbbreviations();
\end{lstlisting}
\end{framed}
\lstinline!Expr! is the main variable type of \csl, internal representation of a symbolic mathematical expression. The output in terminal is the following:
\begin{framed}
\begin{lstlisting}
    <|M|^2> = 1/4*s_14*s_23*(1/2*Ab_0001^(*)*Ab_0002 + Ab_0001*Ab_0001^(*) 
                           + 1/2*Ab_0001*Ab_0002^(*) + 1/4*Ab_0002*Ab_0002^(*))
    Ab_0001 = i*g_L^2/s_13
    Ab_0002 = i*g_L^2/s_12
\end{lstlisting}
\end{framed}
One can see that abbreviations have been introduced by \marty. They can be expanded and the result can be further factored by \csl\ typing 
\begin{framed}
\begin{lstlisting}
    Evaluate(squared_ampl, eval::abbreviation); // Evaluate abbreviations
    DeepFactor(squared_ampl); // Factor the whole expression
    cout << "<|M|^2> = " << squared_ampl << endl;
\end{lstlisting}
\end{framed}
In this way, one can obtain a compact result:
\begin{framed}
\begin{lstlisting}
    <|M|^2> = 1/16*g_L^4*(s_12^(-2) + 4*s_13^(-2) + 4/(s_12*s_13))*s_14*s_23
\end{lstlisting}
\end{framed}
As one can see above, \marty's outputs contain scalar products of external momenta $s_{ij}\equiv p_i\cdot p_j$. Using kinematics this could be further simplified. Considering massless particles and traditional Mandelstam variables for $2\to 2$ process, one has
\begin{align}
    &s_{12} = s_{34} = \frac{1}{2}s,\\
    &s_{13} = s_{24} = -\frac{1}{2}t,\\
    &s_{14} = s_{23} = -\frac{1}{2}u.
\end{align}
\marty\ does not perform any kinematics for now, i.e. stops the simplification as in the output shown. This could be easily implemented in the future. However as it does not represent an important analytical challenge by hand and that it has no impact on the following numerical evaluation (see section \ref{sec:lib}), this part is for now left to the user.

\subsection{Wilson coefficients}

Wilson coefficients are the coefficients in front of particular operator structures in an amplitude \cite{buras}. For the $b\to s\gamma$ process that will be detailed in section \ref{sec:example}, the amplitude may be decomposed on a two operator basis, each one with a scalar coefficient in front. Naming $q$ the photon momentum and $\epsilon$ its polarization vector, one obtains
\begin{equation}
    i\mathcal{M}(b\to s\gamma) = \frac{-4G_F}{\sqrt{2}}\frac{e}{16\pi ^2}V_{tb}V_{ts}^*m_b\left(C_7 \langle\op _7\rangle + C_7^\prime \langle\op_7^\prime\rangle\right),
\end{equation}
with 
\begin{align}
\label{eq:op}
    &\langle\op_7^{(\prime)}\rangle \equiv \langle s\gamma|\op^{(\prime)}_7|b\rangle = \bar{s}\sigma^{\mu\nu}P_{R(L)}bF_{\mu\nu},\\
    &\sigma _{\mu\nu} = \frac{i}{2}\left[\gamma _\mu, \gamma_\nu\right],\\
    &F_{\mu\nu} = iq_{[\mu}\epsilon _{\nu]} = \frac{i}{2}\left(q_{\mu}\epsilon _{\nu} - q_{\nu}\epsilon _{\mu}\right).
\end{align}

The global factor $\frac{-4G_F}{\sqrt{2}}\frac{e}{16\pi ^2}V_{tb}V_{ts}^*m_b$ is defined by convention. 
This procedure to decompose amplitudes in Wilson coefficients and operator matrix elements is used in particular in flavour physics. As quarks appear only in bound states, the partonic amplitude is not the full story. One has to take into account long-distance effects that cannot be calculated pertubatively. The $b\to s\gamma$ transition may correspond for example to a hadronic process $\bar{B^0}\to \bar{K^0}\gamma$. These long-distance effects are model-independent and arise only in the operator matrix element between final and initial states $\langle F|\op|I\rangle$. The BSM dependence is then contained in the Wilson coefficient that can be calculated perturbatively by \marty. 

For now, operators of dimension 6 with 4 fermions cannot be directly given by \marty\ at one-loop as some simplifications are needed that are not yet implemented. The step missing is a double application of Fiertz identities, to simplify all possible momenta in fermion currents. A Wilson coefficient for such an operator could still be read off in the results but would ask the user to do some algebra by hand, to determine which part of the amplitude contributes to the coefficient.

A concrete example of Wilson coefficient calculation is presented in section \ref{sec:example} which is devoted to the calculation of $C_7$ in the MSSM.

\subsection{Library generation}
\label{sec:lib}

Results of \marty\ can in general not be used directly. Depending on the process, it may be a very big and complicated analytical expression. What the user may need are numbers, i.e. numerical evaluations of the analytical results, for a given set of values of the model parameters. Let us consider the cross-section of section \ref{sec:xsec}. The result is rather simple, but the principle would be exactly the same for a more complicated expression. The way it works in \marty\ is also rather simple and is contained in a few lines (giving a library name and a path to create it):
\begin{framed}
\begin{lstlisting}
    mty::Library myLib("uubar_to_ddbar", ".");
    myLib.addFunction("squared_ampl", squared_ampl);
    myLib.build();
\end{lstlisting}
\end{framed}
A \lstinline!mty::Library! is an abstract object that takes symbolic expressions as functions (here the cross-section), creates and compiles a C++ library allowing to evaluate them numerically.
The function generated by \marty\ is:
\begin{framed}
\begin{lstlisting}
    complex_t squared_ampl(
        const complex_t g_L,
        const complex_t s_12,
        const complex_t s_13,
        const complex_t s_14,
        const complex_t s_23
        )
    {     
        return 0.0625*std::pow(g_L, 4)*(std::pow(s_12, -2) + 4*std::pow(s_13, -2) 
                + 4/(s_12*s_13))*s_14*s_23;
    }
\end{lstlisting}
\end{framed}
The function takes as arguments all symbols (possibly complex here) that did not contain any value at the time the library was generated. The library is compiled automatically and can be used as demonstrated in the following.
\begin{framed}
\begin{lstlisting}
    #include "uubar_to_ddbar.h"

    using namespace std;
    using namespace uubar_to_ddbar;

    int main() {

        cout << "XSec = " << squared_ampl(0.1, 100, 60, 40, 40) << endl;
        return 0;
    }
\end{lstlisting}
\end{framed}
A library may contain as many functions as wanted. This procedure is fully general and is automated. Note that if the library needs additional include or library paths (in particular if \csl\ and \marty\ are not installed in standard locations), it is possible to specify them with:
\begin{framed}
\begin{lstlisting}
    myLib.addIPath("/home/.local/include");
    myLib.addLPath("/home/.local/lib");
\end{lstlisting}
\end{framed}

\subsection{Feynman diagrams}

\begin{figure}[!h]
    \centering
    \includegraphics[width=0.85\linewidth]{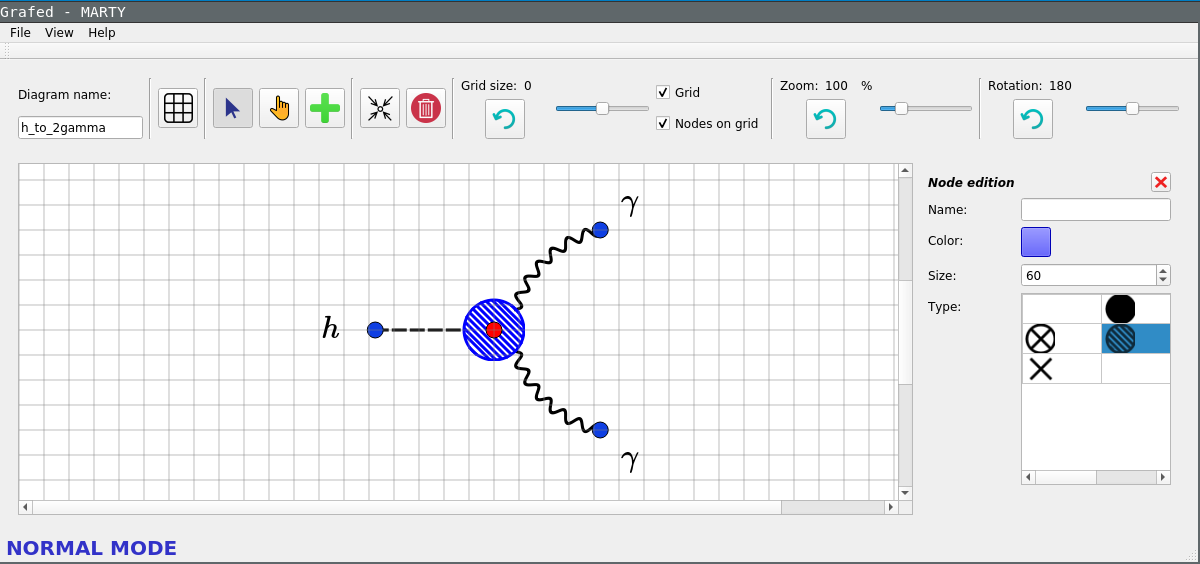}
    \caption{Generic $h\to\gamma\gamma$ diagram creation with \grafed. Node colour / type / name / size, edge colour / type / curve / thickness can be edited easily with the application to create any kind of diagram. Diagrams may then be exported in a .png file directly.}
    \label{fig:my_label}
\end{figure}

\grafed\ is the part of \marty\ generating and rendering Feynman diagrams. It is used to create automatically diagrams when calculating a process with the \lstinline!Show(res)! command as we discussed in section \ref{sec:ampl}. It can also be used to edit or create diagrams from scratch. Many aspects of the diagrams can be chosen by the user in an intuitive way. A screenshot of \grafed\ is shown in figure \ref{fig:my_label}.

\grafed\ will in the future be released as standalone. All diagrams in this publication are generated automatically or created with \grafed.

\section{Calculation of $\delta ^{LO}C_7^{\chi,\tilde{t}}(M_W)$ in the pMSSM}
\label{sec:example}
An example of \marty's capabilities is presented in this section, namely the calculation of the MSSM contribution to the Wilson coefficient $C_7$. This coefficient describes the $b\to s\gamma$ transition and is non zero only at the one-loop level. 

\subsection{The pMSSM}
We consider here the phenomenological Minimal Supersymmetric Standard Model (pMSSM) which is a generic CP conserving MSSM framework that has 19 parameters more than in the SM as opposed to the full MSSM which has 105 extra parameters. The pMSSM has been chosen for validation because of its complexity and generality. Obtaining a correct result in this model demonstrates \marty's capabilities for model building and symbolic calculations.

\subsection{The Wilson coefficient $C_7$}

\begin{figure}[!h]
    \centering
    \includegraphics[width=0.3\linewidth]{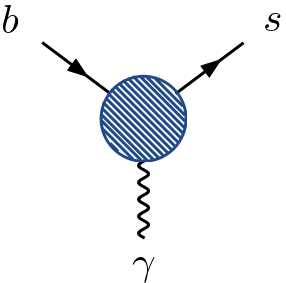}
    \caption{$b\to s\gamma$ process represented in a model-independent way. The transition amplitude is the sum of all diagrams that can fill correctly the hatched disk. This diagram has been built using \emph{GRAFED}.}
    \label{fig:bsgamma}
\end{figure}

We calculate the Leading Order (LO) value of the Wilson coefficient $C_7$, associated with the operator in equation \ref{eq:op}. The process is showed in figure \ref{fig:bsgamma}. It is a FCNC process with a photon changing the quark flavour $b\to s$ which is forbidden at tree-level in the SM and the pMSSM, and the LO is thus at the one-loop level.
Strong experimental constraints exist for FCNCs and their calculations for BSM models represent an important task for phenomenology. 

\begin{figure}
    \centering
    \includegraphics[width=0.3\linewidth]{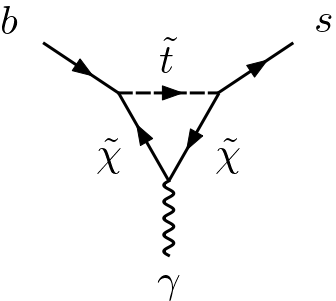}
    \hspace{2.5cm}
    \includegraphics[width=0.3\linewidth]{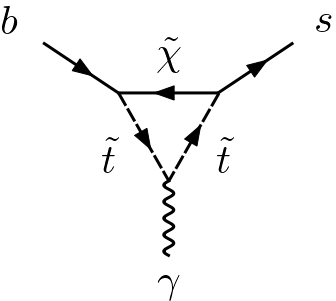}
    \caption{Two types of contribution for $C_7$ in the pMSSM, with stops $\tilde{t}$ and charginos $\tilde{\chi}$. These diagrams have been built using \emph{GRAFED}.}
    \label{fig:c7susy}
\end{figure}

We consider in this example one of the supersymmetric contributions i.e. diagrams with the top squarks and charginos shown in figure \ref{fig:c7susy}. We perform the calculation on-shell in the Feynman-'t Hooft gauge\footnote{Other gauges can be used such as the unitary or Lorentz gauges.}. The reversal of the fermion-flow in the diagram is due to fermion-number violating interactions between charginos and SM fermions. This is mostly related to the definition of charginos and may be treated following prescriptions of \cite{majoranarules}. At the end of the calculation, the fermion flow is regular but may get a sign due to charge conjugation matrix $C$ which appears. This sign is important to determine exactly because of interferences between the diagrams.

We vary two pMSSM parameters, $\mu$ (the Higgsino parameter) and $M_2$ (the Wino mass). More details on MSSM parameters are given in \cite{susyprimer}. Contributions to $C_7$ come from chargino and stop loops and depend on $\mu$ and $M_2$ in particular. The chargino mass matrix reads 
\begin{equation}
    M_{\chi} = \left(\begin{array}{cc}
        0 & X^T \\
        X & 0 
    \end{array}\right),
\end{equation}
with 
\begin{equation}
    X = \left(\begin{array}{cc}
        M_2 & \sqrt{2}\sin \beta M_W \\
        \sqrt{2}\cos \beta M_W & \mu 
    \end{array}\right),
\end{equation}
$\tan\beta$ being the angle between the two Higgs doublets' Vacuum Expectation Values (VEVs). 

The stop squared mass matrix reads
\begin{equation}
    M^2_{\tilde{t}} = \left(\begin{array}{cc}
       m_{Q_3}^2 + m_t^2 + \Delta _{\tilde{u}_L}  & v(A_t^*\sin\beta - \mu y_t\cos\beta) \\
        v(A_t\sin\beta - \mu^*y_t\cos\beta)  & m_{u_3}^2 + m_t^2 + \Delta _{\tilde{u}_R}
    \end{array}\right).
\end{equation}
$m_{Q_3}$, $m_{u_3}$ are soft supersymmetry breaking parameters, $A_t$ is a trilinear coupling, $y_t$ the top Yukawa and finally 
\begin{align}
    \Delta _{\tilde{u}_L} &= \left(\frac{1}{2} - \frac{2}{3}\sin^2 \theta _W\right)\cos(2\beta)M^2_Z,\\
    \Delta _{\tilde{u}_R} &= \frac{2}{3}\sin^2\theta_W\cos(2\beta)M^2_Z.
\end{align}

The exact numerical values of pMSSM parameters used to evaluate $C_7$ are presented in table \ref{tab:num}.

\begin{table}[!htbp]
    \centering
    \begin{tabular}{c|c}
         Parameter & Value  \\\hline
         $A_t$ & 500\\
         $m_{Q_3}$ & 1000 GeV\\
         $m_{u_3}$ & 1000 GeV\\
         $\tan\beta$ & 50\\
         $\mu$ & $[-800, 800]\text{ GeV}$\\
         $M_2$ & $[-1000, 1000]\text{ GeV}$
    \end{tabular}
    \caption{Numerical values of supersymmetric parameters used to evaluate $C_7$. $M_2$ and $\mu$ are varied in the given ranges. Other pMSSM parameters are irrelevant for the calculation presented here.}
    \label{tab:num}
\end{table}

\begin{figure}[!h]
    \centering
    \includegraphics[width=0.43\linewidth]{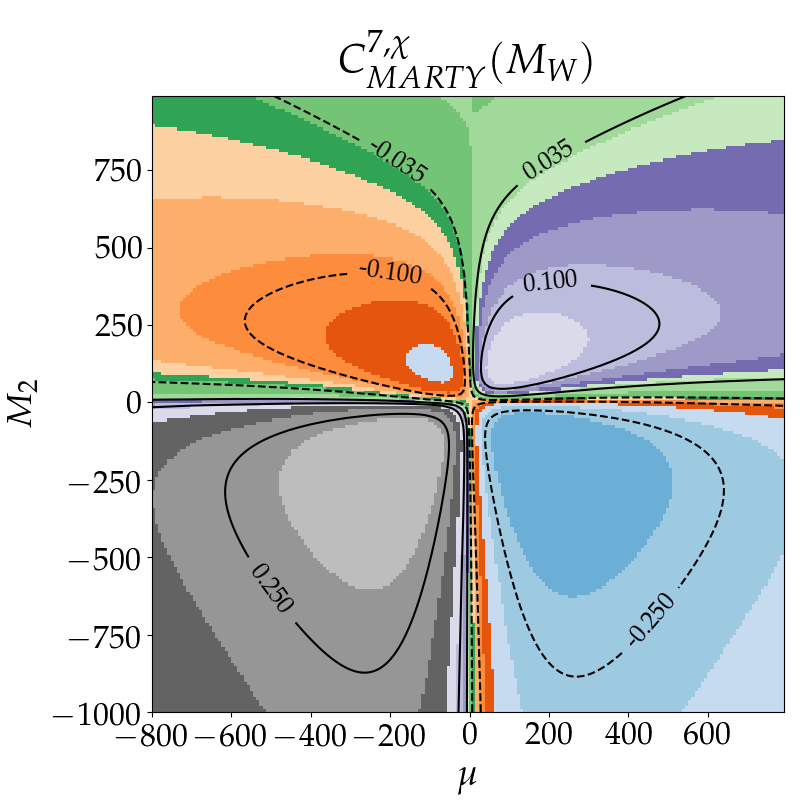}
    \includegraphics[width=0.43\linewidth]{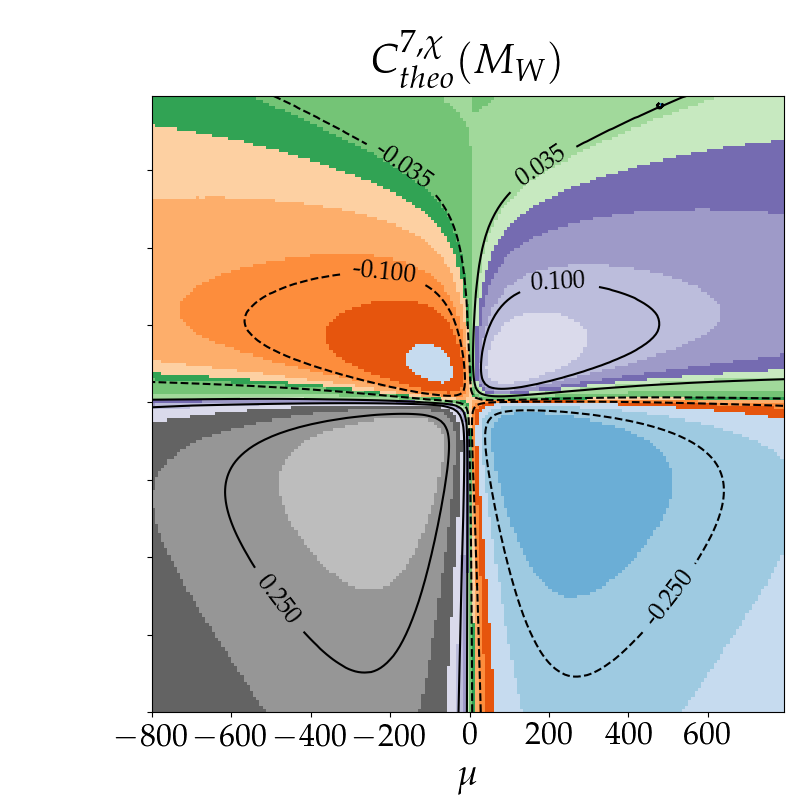}
    \raisebox{0.59cm}{\includegraphics[width=0.0995\linewidth]{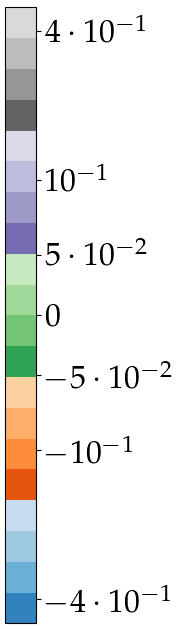}}
    \caption{Results for $C_7$ (chargino and stop contributions) in the pMSSM, from \marty\ one the left and from the analytical formula \cite{C7_charg} on the right, for the spectrum generated by \marty at tree-level. The results match to four digits in average.}
    \label{fig:res}
\end{figure}

\begin{figure}[!h]
    \centering
    \includegraphics[width=0.43\linewidth]{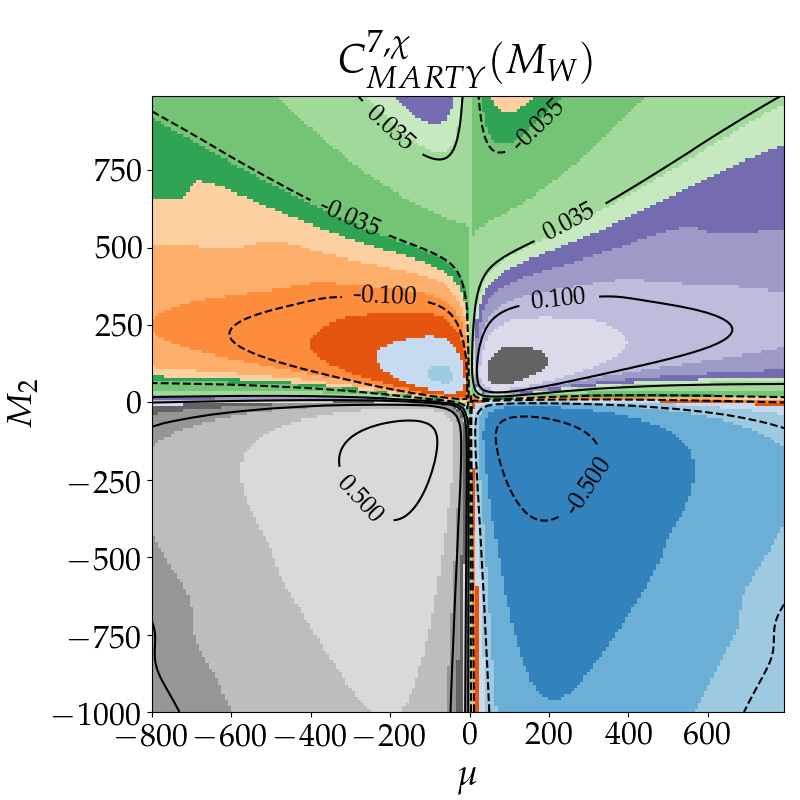}
    \includegraphics[width=0.43\linewidth]{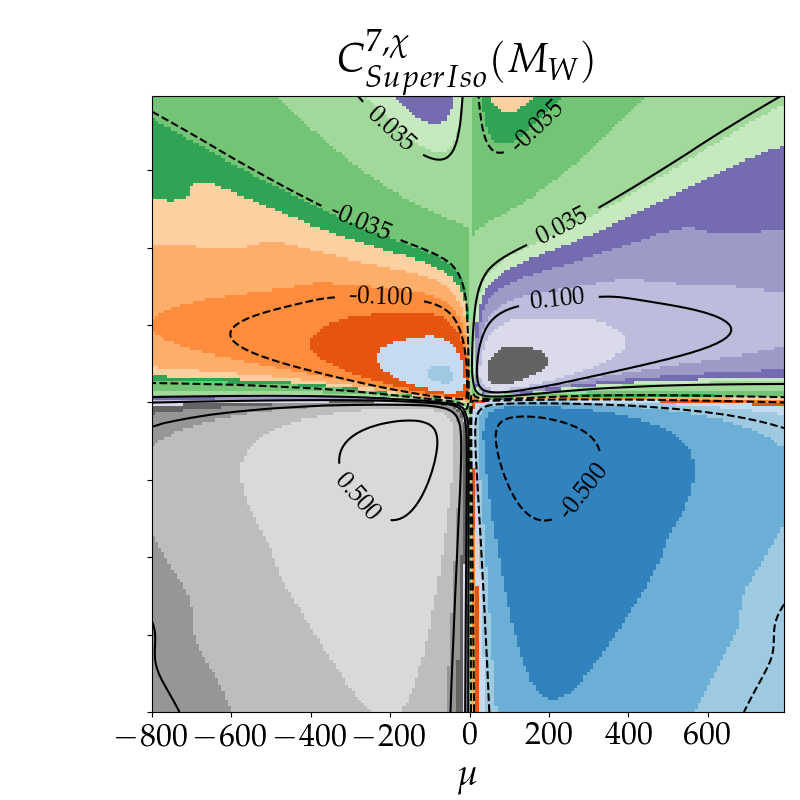}
    \raisebox{0.59cm}{\includegraphics[width=0.0995\linewidth]{colorbar.png}}
    \caption{Results for $C_7$ (chargino and stop contributions) in the pMSSM, from \marty\ one the left and from the output of SuperIso \cite{superiso} on the right, for the spectrum generated by SOFTSUSY  \cite{softsusy1, softsusy2} with two-loop corrections. The results match to four digits in average.}
    \label{fig:res2}
\end{figure}

The results are shown in figures \ref{fig:res} and \ref{fig:res2}. \marty's output is compared with the analytical formula given in \cite{C7_charg} and with SuperIso \cite{Mahmoudi:2007vz,superiso,Mahmoudi:2009zz}. Numerical evaluations have been done for two different spectra. This first one (figure \ref{fig:res}) is a tree-level spectrum computed by \marty\ using GSL \cite{gsl}, and the result is compared with the analytical formula. The second spectrum (figure \ref{fig:res2}) is calculated by SOFTSUSY \cite{softsusy1,softsusy2} with two-loop order corrections which are known to be important for the charginos \cite{susyprimer}. For this spectrum, we compare \marty\ with the output of SuperIso.

As can be seen for all the results there exists an excellent agreement between the analytical formula, SuperIso, and \marty. The agreement is up to 4 digits. In addition, we also tested the results given by FormCalc \cite{formcalc} for the same process, and there is in this case a perfect agreement with \marty\ with 10 identical digits in average, for both spectra. One explanation may be that we used quadruple precision (128 bits) floating point variables for FormCalc and \marty's outputs, and only double precision (64 bits) for SuperIso output and the analytical formula. The 4-digits precision is however completely satisfactory considering the uncertainty coming form higher orders in perturbation theory.

This example completes the presentation of what \marty\ can calculate. \marty\ will generate for any process, in any model, libraries evaluating theoretical quantities and give the user a spectrum generator at the same time. In the case of supersymmetry, spectrum generators already exist with higher-order terms but in a general BSM models one needs this generic tree-level spectrum generator. More information and example can be found on the website \url{https://marty.in2p3.fr}. 

\section{Performance}

We measure the performance of a computer program with two main indicators. The execution speed and the quantity of memory (RAM) the program needs for running. For BSM symbolic calculations at one-loop, it is not possible to give a standard execution speed nor the quantity of memory as it depends on the model and the process to calculate. The amount of memory taken by \marty\ is typically very small. It is very rare to reach 1~GB, and is often under 100~MB. Indicative values of execution times are shown in table \ref{tab:perf}, measured on various processes, always running on a single CPU. 

\begin{table}[!htbp]
    \centering
    \begin{tabular}{c|c c}
         External legs & Tree-level & One-loop  \\\hline
         2 & $ \leq 10^{-1}$ & $ 10^{-1}$\\
         3 & $ \leq 10^{-1}$ & $ 10^{0}/10^1 $\\
         4 & $ \leq 10^{-1}$ & $ 10^{2} $\\
    \end{tabular}
    \caption{Typical execution time (order of magnitude) of \marty\ in second per 100 Feynman diagrams during the calculation of an amplitude, for different numbers of external legs.}
    \label{tab:perf}
\end{table}

It can be seen in table \ref{tab:perf} that the calculation complexity depends strongly on the number of legs at one-loop. The more legs there is for the loop, the more terms appear in the amplitude. This number of terms grows very fast with the number of legs connected to the loop and explains the results shown here.

For squared amplitudes there is no simple rule to determine the execution time but it is in general several orders of magnitude more than the simple amplitude calculation as squaring the amplitude also squares the number of terms to simplify. Improving performance for this calculation is an important development for the next release of \marty.

\section{Future developments}

\marty\ has fulfilled most of the planned requirements, but further developments are ongoing which are listed in the following.
\begin{itemize}
    \item \textbf{Wilson coefficients for 4-fermion operators.} For now, \marty\ can compute amplitudes for 4-fermions processes, but cannot give automatically the corresponding Wilson coefficients because of a missing simplification step. This step is the double application of Fiertz identities to simplify all momenta in quark currents. Once this simplification will be implemented, 4-quarks operators will become available in \marty.
    \item \textbf{Faster and lighter calculation of squared amplitude.} Squared amplitudes are heavy to compute (typically $N^2$ terms for an amplitude with $N$ terms). At the one-loop order, the calculation of squared amplitudes is currently very heavy, and future optimizations are required.
    \item \textbf{More group theory simplifications.} All simplifications with algebra generators are not implemented in \marty. Some are missing because it is very difficult to automate these identities for all semi-simple groups and all representations. These missing simplifications concern mostly non-fundamental representations, exceptional algebras and squared amplitudes for pure gluonic amplitudes. Further developments will focus on this issue but the user can also define easily the missing properties.
    \item \textbf{No automated NLO corrections}. With \marty\ one can calculate all one-loop quantities needed to renormalize a BSM model. However this procedure is not automated and will surely be a point of attention in the future.
    \item \textbf{Operator mixing for Wilson coefficients.} Renormalization comes with operator mixing for Wilson coefficients. This task is more challenging but there is currently no code able to fully automate this procedure for general BSM. Therefore having a code able to do this task would be very useful for flavour physics.
    \item \textbf{Interfaces with other codes.} This publication presents the first version of \marty, that is at the moment not interfaced with other codes. This is the most important improvement that is planned for next developments. Universal Feynman Rules Output (UFO \cite{ufo}) will link \marty\ directly to event generators, and work is in progress to have direct interface with SuperIso \cite{Mahmoudi:2007vz,superiso,Mahmoudi:2009zz} for flavour physics, and SuperIso Relic \cite{Arbey:2009gu,Arbey:2011zz,Arbey:2018msw} for dark matter phenomenology.
\end{itemize}

\section{Conclusion and Outlook}

We presented \marty, a new C++ framework automating theoretical calculations symbolically for BSM physics. The degree of generality reached by \marty\ has never been achieved before. It has its own computer algebra system (\csl) and automates all theoretical calculations directly from the Lagrangian. Feynman rules, Feynman diagrams, amplitudes, cross-sections, and Wilson coefficients can be obtained in a very large variety of BSM models up to the one-loop level. A full NLO treatment will also be implemented in the near future, treating renormalization of fields, masses, couplings, and Wilson coefficients (including operator mixings).

A proof of its capabilities has been demonstrated through a tree-level cross-section calculation, and a one-loop Wilson coefficient in the pMSSM. The results are at first symbolic mathematical expressions, but numerical C++ libraries are built automatically by \marty\ allowing us to explore in full generality the parameter space of the model for some user-defined quantities. A spectrum generator specific to the user's model is also created automatically by \marty\ when needed. Most of popular BSM models can be built in \marty. The MSSM, extended gauge models, and vector-like quarks are examples of possible BSM implementations. 

\marty\ can already be very useful for BSM phenomenology in this current version. The particular advantage of \marty\ is to be written as a unique code, not depending on any framework. Within \marty, every aspect of model building and high-energy physics calculation are under control, in the same program and in the same language. This is a unique opportunity for future collaborations to take this code even further, extending it to new models, other simplification methods, or even different types of calculations.

\bibliographystyle{elsarticle-num}
\bibliography{article}

\end{document}